\documentclass[aps,showpacs,twocolumn]{revtex4}
\usepackage{epsfig}
\usepackage{graphicx,color,dcolumn}
\usepackage{epstopdf}
\usepackage{amsmath,amssymb}
\usepackage{multirow}
\usepackage{changes}
\usepackage{booktabs}
\usepackage{threeparttable}

\renewcommand{\arraystretch}{1.5} 

\begin{document}

\title{Strange hidden-charm tetraquarks in constituent quark models}

\author{Xin Jin$^1$}\email[E-mail: ]{181002005@stu.njnu.edu.cn}
\author{Xuejie Liu$^2$}\email[E-mail: ]{1830592517@qq.com}
\author{Yaoyao Xue$^1$}\email[E-mail: ]{181002022@stu.njnu.edu.cn}
\author{Hongxia Huang$^1$}
\email[E-mail: ]{hxhuang@njnu.edu.cn (Corresponding author)}
\author{Jialun Ping$^1$}
\email[E-mail: ]{jlping@njnu.edu.cn (Corresponding author)}
\affiliation{$^1$Department of Physics, Nanjing Normal University, Nanjing 210023, P.R. China}
\affiliation{$^2$School of Physics, Southeast University, Nanjing 210094, P. R. China}

\begin{abstract}
Inspired by the newly reported $Z_{cs}(3985)^{-}$ by the BESIII Collaboration, we systematically investigate the strange hidden-charm tetraquark systems $cs\bar{c}\bar{u}$ with two structures: meson-meson and diquark-antidiquark. Two quark models: the chiral quark model (ChQM) and the quark delocalization color screening model (QDCSM) are used here. Similar results are obtained in both two quark models. There is no any bound state in either ChQM or QDCSM, which excludes the molecular state explanation ($D_{s}D^{*}/D_{s}^{*}D/D_{s}^{*}D^{*}$) of the reported $Z_{cs}(3985)^{-}$. However, the effective potentials for the diquark-antidiquark $cs\bar{c}\bar{u}$ systems shows the possibility of some resonance states with mass range of $3916.5\sim 3964.6$ MeV for $IJ^{P}=\frac{1}{2} 0^{+}$, $4008.8\sim 4091.2$ MeV for $IJ^{P}=\frac{1}{2} 1^{+}$, $4246.8\sim 4418.1$ MeV for $IJ^{P}=\frac{1}{2} 2^{+}$. So the observed $Z_{cs}(3985)^{-}$ state is possible to be explained as a compact resonance state composed of $cs\bar{c}\bar{u}$ with $IJ^{P}=\frac{1}{2} 0^{+}$ or $IJ^{P}=\frac{1}{2} 1^{+}$. The study of the scattering process of corresponding open channels is under way to check this conclusion.
\end{abstract}


\maketitle

\section{Introduction} \label{introduction}
In 2013, the BES\uppercase\expandafter{\romannumeral3} Collaboration studied the $e^{+}e^{-}\rightarrow  J/\psi \pi^{+}\pi^{-}$ process at a center-of-mass energy of 4.26 GeV and reported a new charged charmonium-like structure in the $\pi^{\pm} J/\psi $ invariant spectrum, which is called $Z_{c}(3900)$ and has a mass of $3899.0\pm3.6\pm4.9$ MeV and a width of $46\pm10\pm20$ MeV~\cite{Ablikim:2013mio}. At the same time, the Belle observed a $Z_{c}(3895)\pm$ state, with a mass of $3894.5\pm6.6\pm4.5$ MeV and a width of $63\pm24\pm26$ MeV in the process $Y(4260)\rightarrow J/\psi \pi^{+}\pi^{-}$~\cite{Liu:2013dau}. The mass and width of $Z_{c}(3900)$ and $Z_{c}(3895)\pm$ are very close within errors, so they are the same state~\cite{Liu:2013nqa}. The state $Z_{c}(3900)$ has been further confirmed by the CLEO-c Collaboration in the the decay $Y(4160)\rightarrow J/\psi \pi^{+}\pi^{-}$~\cite{Xiao:2013iha}. Subsequently, there is a lot of theoretical work to study $Z_{c}(3900)$ and its associated states. Some researches treated it as the tightly bound diquark-antidiquark state~\cite{Dias:2013xfa,Wang:2013vex,Deng:2014gqa,Agaev:2016dev,Liu:2019gmh,Chen:2019osl}. However, many researches treated it as molecular states~\cite{Wilbring:2013cha,Dong:2013iqa,Gutsche:2014zda,Chen:2015igx,Esposito:2014hsa,Gong:2016hlt,Ke:2016owt}. Some indicated that the $Z_{c}(3900)$ is not a usual resonance but a threshold cusp~\cite{Swanson:2014tra,Ikeda:2016zwx}. Lattice QCD method is also used to find $Z_{c}(3900)$, but it has not been found~\cite{Prelovsek:2013xba,Prelovsek:2013sxa,Chen:2014afa,Prelovsek:2014swa}.


Very recently, the BESIII Collaboration reported their study of the processes of $e^{+}e^{-} \rightarrow K^{+}(D_{s}^{-}D^{*0}+D_{s}^{*-}D^{0})$, and found a new structure $Z_{cs}(3985)^{-}$ near the $D_{s}^{-}D^{*0}/D_{s}^{*-}D^{0}$ thresholds in the $K^{+}$ recoil-mass spectrum for events collected at $\sqrt{s}=4.681$ GeV. The pole mass and width of this state are ($3982.5^{+1.8}_{-2.6}\pm2.1$) Mev and ($12.8^{+5.3}_{-4.4}\pm3.0$) Mev, respectively~\cite{Ablikim:2020hsk}. From the production mode, it is easy to get that the minimum quark component of $Z_{cs}(3985)^{-}$ is $c\bar{c}s\bar{q}$ ($q=u/d$), and this state should be a partner structure of the well-known $Z_{c}(3885)^{-}$ reported in $e^{+}e^{-} \rightarrow D^{*-}D^{0}\pi^{+}$~\cite{Ablikim:2014}. Besides, it is the first candidate of the charged
hidden-charm tetraquark state with strangeness, whose discovery can provide more hints to the quest of charged exotic $Z$ structures. Therefore, the observation of $Z_{cs}(3985)^{-}$ immediately stimulated a lot of theoretical discussions~\cite{Wang:2020kej,Meng:2020ihj,Liu:2020nge,Yang:2020nrt,Chen:2020yvq,Du:2020vwb,Cao:2020cfx,Sun:2020hjw,Rossi:2020ezg,Wang:2020rcx}.

Actually, theoretical predictions of the charged hidden-charm tetraquark with strangeness have been made in different models~\cite{Lee:2009,Voloshin:2019,Ferretti:2020}. D.Ebert used relativistic quark model based on the quasipotential approach to calculate the mass spectra of tetraquarks $[Qs][\bar{Q}\bar{q}]/[Qq][\bar{Q}\bar{s}](Q=c,b)$~\cite{Ebert:2005nc}. They found that all $S-$wave tetraquarks with hidden bottom lie considerably below open bottom thresholds and they should be narrow states which can be observed experimentally. However hidden-charm tetraquark states all above open charm thresholds. Dianyong Chen indicated that there exist enhancement structures with both hidden-charm and open-strange decays, which are near the $D\bar{D^{*}_{s}}/D^{*}\bar{D_{s}}$ and $D^{*}\bar{D^{*}_{s}}/\bar{D^{*}}D^{*}_{s}$ thresholds under the initial single chiral particle emission (ISChE) mechanism~\cite{Chen:2013wca}. Chengrong Deng studied the same charged tetraquark states using the variational method GEM in the color flux-tube model with a four-body confinement potential. The numerical results indicated that some compact resonance states can be formed~\cite{Deng:2014gqa}.

Strong interaction is the strongest of the four interactions in nature, but understanding its nature has always been a difficult problem in physics. Quantum chromodynamics (QCD) is widely accepted as the basic theory of strong interactions. QCD can deal with scattering problems by perturbation expansion in high energy regions, but the spontaneous chiral symmetry breaking and color confinement appear in low energy regions. To study hadron-hadron interactions and multiquark states, many quark models based on QCD theory have been developed. The chiral quark model (ChQM) is one of the accepted models~\cite{Salamanca1}. The interactions in this model include the colorful one-gluon-exchange and confinement, colorless Goldstone boson exchange and the chiral partner $\sigma$ meson-exchange. Another approach is the quark delocalization color screening model (QDCSM), which was developed to study the similarities between nuclear and molecular forces~\cite{Wang:1992wi}. Both of these two models describe the properties of deuteron, nucleon-nucleon and hyperon-nucleon
interactions well~\cite{ChenLZ,ChenM}.
Very recently, both ChQM and QDCSM have been used to explain the full-heavy tetraquark states~\cite{Jin:2020jfc} and the reported $X(6900)$ by LHCb collaboration~\cite{Aaij:2020fnh} could be explained as a compact resonance state in both two models. It is quite natural to extend the study to the charged hidden-charm tetraquark systems with strangeness. So we investigate the tetraquark systems composed of $c\bar{c}s\bar{q}$ ($q=u/d$) in both ChQM and QDCSM in present work.

The structure of this paper is as follows. section II gives a brief introduction of two quark models, and the construction of wave functions.
The numerical results and discussions are given in Section III. The summary is presented in the last section.

\section{Models and wavefunctions}
In this work, we investigate the charged charmonium-like tetraquarks with hidden-charm and open-strange $c\bar{c}s\bar{q}$ ($q=u/d$) within two quark
models: ChQM and QDCSM. Two structures: meson-meson
and diquark-antidiquark, are considered. In this sector, we will introduce
these two models and the wave functions of the tetraquarks
for two structures.

\subsection{The chiral quark model (ChQM)}
The ChQM has been successfully applied to describe the properties of hadrons
and hadron-hadron interactions~\cite{Salamanca1,Salamanca2}. The model details can
be found in Ref.~\cite{Salamanca1,Salamanca2}. We only show
the Hamiltonian of the model here.
\begin{align}
H={}&\sum_{i=1}^{4}( m_i+\frac{p_i^2}{2m_i})-T_{cm}+\sum_{i=1<j}^{4}(
V^{CON}_{ij}+V^{OGE}_{ij})
\end{align}
where $T_{cm}$ is the kinetic energy of the center of mass; $V^{CON}_{ij}$ and
$V^{OGE}_{ij}$ are the interactions of the confinement and the
one-gluon-exchange, respectively. For the $c\bar{c}s\bar{u}$ system, there is no
$\sigma$-exchange interactions, because the $\sigma$ meson cannot be exchanged between $u/d$ quark and $s/c$ quark. The forms of $V^{CON}_{ij}$
and $V^{OGE}_{ij}$ are shown below:
\begin{align}
V^{CON}_{ij}&= -\boldsymbol{\lambda}^c_i \cdot \boldsymbol{\lambda}^c_j (a_c
r^2_{ij}+V_{0_{ij}}) \\
V^{OGE}_{ij}&=\frac{\alpha_{s_{ij}}}{4} \boldsymbol{\lambda}^c_i \cdot
\boldsymbol{\lambda}^c_j
  [\frac{1}{r_{ij}}-\frac{\pi}{2}\delta(\mathbf{r}_{ij})(\frac{1}{m_{i}^{2}}+\frac{1}{m_{j}^{2}}
  +\frac{4\boldsymbol{\sigma}_{i}\cdot\boldsymbol{\sigma}_{j}}{3m_{i}m_{j}})\notag\\
              {}&-\frac{3}{4m_{i}m_{j}r_{ij}^{3}}S_{ij}]\\
S_{ij}&=\{3\frac{(\boldsymbol{\sigma}_{i}\cdot
\boldsymbol{r}_{ij})(\boldsymbol{\sigma}_{j}\cdot
\boldsymbol{r}_{ij})}{r_{ij}^{2}}
  -\boldsymbol{\sigma}_{i}\cdot \boldsymbol{\sigma}_{j}\}
\end{align}
where $S_{ij}$ is quark tensor operator; $\alpha_{s_{ij}}$ is the quark-gluon
coupling constant.

\subsection{The quark delocalization color screening model (QDCSM)}
Generally, the Hamiltonian of QDCSM is almost the same as that of ChQM, but
with two modifications~\cite{Wang:1992wi}. The one is that there
is no $\sigma$-meson exchange in QDCSM, and another one is that the screened
color confinement is used between quark pairs reside in different
clusters, aiming to take into account the QCD effect which has not yet been
included in the two-body confinement. The confining potential in QDCSM was
modified as follows:
 $$V^{CON}_{ij}=\left\{
 \begin{array}{rcl}
 -\boldsymbol{\lambda}^c_i \cdot \boldsymbol{\lambda}^c_j (a_c
 r^2_{ij}+V_{0_{ij}})&&\text{i,j in the same cluster}\\
 - \boldsymbol{\lambda}^c_i \cdot \boldsymbol{\lambda}^c_j a_c\frac{1 -
 e^{-\mu_{ij} r^2_{ij}}}{\mu_{ij}}&&\text{, otherwise}\\
 \end{array}\right.$$
where $\mu_{ij}$ is the color screening parameter, which is determined by
fitting the deuteron properties, $NN$ scattering phase shifts, and
$N\Lambda$ and $N\Sigma$ scattering phase shifts, respectively, with
$\mu_{uu}=0.45~$fm$^{-2}$, $\mu_{us}=0.19~$fm$^{-2}$ and
$\mu_{ss}=0.08~$fm$^{-2}$, satisfying the relation,
$\mu_{us}^{2}=\mu_{uu}\mu_{ss}=0.19~$fm$^{-2}$~\cite{ChenM}. When extending to the heavy quark
case, there is no experimental
data available, so we take it as a adjustable parameter $\mu_{cc}=0.01 \sim
0.001~$fm$^{-2}$ and we find the results are insensitive to the value of $\mu_{cc}$. So in the present
work, we take $\mu_{cc}=0.01~$fm$^{-2}$, $\mu_{uc}=0.067~$fm$^{-2}$ and $\mu_{sc}=0.0283~$fm$^{-2}$.

The single particle orbital wave functions in the ordinary quark cluster model
are the left and right centered single Gaussian functions:
\begin{eqnarray}
\phi_\alpha(\mathbf{S}_{i})=\left(\frac{1}{\pi
b^2}\right)^{\frac{3}{4}}e^ {-\frac{(\mathbf{r}-\mathbf{S}_i/2)^2}{2b^2}},
 \nonumber\\
\phi_\beta(-\mathbf{S}_{i})=\left(\frac{1}{\pi
b^2}\right)^{\frac{3}{4}}e^ {-\frac{(\mathbf{r}+\mathbf{S}_i/2)^2}{2b^2}}.
\end{eqnarray}
The quark delocalization in QDCSM is realized by writing the single particle
orbital wave function as a linear combination of the left and right Gaussians:
\begin{eqnarray}
{\psi}_{\alpha}(\mathbf{S}_{i},\epsilon) &=&
\left({\phi}_{\alpha}(\mathbf{S}_{i})
+\epsilon{\phi}_{\alpha}(-\mathbf{S}_{i})\right)/N(\epsilon),
\nonumber \\
{\psi}_{\beta}(-\mathbf{S}_{i},\epsilon) &=&
\left({\phi}_{\beta}(-\mathbf{S}_{i})
+\epsilon{\phi}_{\beta}(\mathbf{S}_{i})\right)/N(\epsilon),
\nonumber \\
N(\epsilon)&=&\sqrt{1+\epsilon^2+2\epsilon e^{{-S}_i^2/4b^2}}.
\end{eqnarray}
where $\epsilon(\mathbf{S}_i)$ is the delocalization parameter determined by
the dynamics of the quark system rather than
adjusted parameters. In this way, the system can choose its most favorable
configuration through its own dynamics in a larger
Hilbert space.

The parameters used in our previous work are determined by fitting the mass
spectrum of mesons and baryons including light
quarks ($u,~d,~s$), but the mesons composed of heavy quarks like
$\eta_{b}$($\eta_{c}$) or $\Upsilon$($J/\psi$) do not fit well.
To give the right mass of the mesons we used in this work, we adjust the
parameters by fitting the masses of mesons that we need. Since there is no $\sigma$-exchange interactions in ChQM for the $c\bar{c}s\bar{u}$ system and the delocalization and color screening in QDCSM work in quarks of different clusters, there is no discrepancy of these two models in describing mesons. So the parameters of two quark models are the same, which are shown in Table~\ref{parameters}. The calculated masses of the mesons
are shown in Table~\ref{mesons}.
\begin{table}[!htb]
\begin{center}
\caption{Model parameters.}
\renewcommand\arraystretch{1.8}

\begin{tabular}{ccc}
\hline
 \hline
State & $b$ (fm)& 0.3 \\
      & $m_u$ (MeV) & 313 \\
      & $m_s$ (MeV) & 536 \\
      & $m_c$ (MeV) & 1728 \\
 \hline
Confinement & $a_{c}$ (MeV fm$^{-2}$) & 101 \\
            & $V_{0_{us}}$ (MeV) & -180.6 \\
            & $V_{0_{uc}}$ (MeV) & -133.6 \\
            & $V_{0_{sc}}$ (MeV) & -68.1 \\
            & $V_{0_{cc}}$ (MeV) & 76 \\
 \hline
OGE & $\alpha_{s_{us}}$ & 0.33 \\
    & $\alpha_{s_{uc}}$ & 0.38 \\
    & $\alpha_{s_{uc}}$ & 0.66 \\
    & $\alpha_{s_{uc}}$ & 1.67 \\
    \hline
\end{tabular}
\label{parameters}
 \end{center}
 \end{table}

 \begin{table}[!htb]
\begin{center}
\caption{The caculated masses (in MeV) of the mesons. Experimental values are taken from the Particle Data Group(PDG).}
\begin{tabular}{cccccccccc}
\hline
 \hline
    & K & $K^{*}$ & $\eta_{c}$ & $J/\psi$ & $D$ & $D^{*}$ & $D_{s}$ & $D_{s}^{*}$ \\
 \hline
  Exp.  & 494 & 892 & 2984 & 3097 & 1865 & 2007 & 1968 & 2112 \\
  Model & 495 & 892 & 2984 & 3097 & 1865 & 2007 & 1968 & 2112 \\
 \hline
  \hline
  \end{tabular}
\label{mesons}
 \end{center}
 \end{table}

\subsection{The wave function}
In this work, we study a bound-state problem by a well-established method, the resonating group method (RGM)~\cite{RGM}. In order to calculate the energy of the $c\bar{c}s\bar{u}$ system, we first constructed the four-quark wave function in the following form:
\begin{equation}
\Psi=\mathcal{A}[[\psi^{L}\chi^{\sigma}]_{JM}\chi^{f}\chi^{c}].
\end{equation}
where $\psi^{L},\chi^{\sigma},\chi^{f}$ and $\chi^{c}$ are the orbital, spin,
flavor and color wave functions, respectively,
which are given below. The symbol $\mathcal{A}$ is the anti-symmetrization
operator. For the $c\bar{c}s\bar{u}$ system, $\mathcal{A}=1$ because the quarks are not identical particles in the $SU(3)$ symmetry and no anti-symmetrization requirement is needed here.

\subsubsection{The orbital wave function}
The total orbital wave function is composed of two internal cluster orbital
wave functions ($\psi_{1}(\mathbf{R}_{1})$ and
$\psi_{2}(\mathbf{R}_{2})$), and one relative motion wave function
($\chi_{L}(\mathbf{R})$) between two clusters.
\begin{equation}
\psi^{L}=\psi_{1}(\mathbf{R}_{1})\psi_{2}(\mathbf{R}_{2})\chi_{L}(\mathbf{R})
\end{equation}
where $\mathbf{R}_{1}$ and $\mathbf{R}_{2}$ are the internal coordinates for
the cluster 1 and cluster 2, respectively.
$\mathbf{R}=\mathbf{R}_{1}-\mathbf{R}_{2}$ is the relative coordinate between
the two clusters 1 and 2. $\chi_{L}(\mathbf{R})$ is expanded by gaussian bases
\begin{align}
\chi_{L}(\mathbf{R})={}&\frac{1}{\sqrt{4\pi}}(\frac{3}{2\pi
b^{2}})\sum_{i=1}^{n}C_{i}\notag\\
                     {}&\times\int \exp
                     [-\frac{3}{4b^{2}}(\mathbf{R}-\mathbf{s}_{i})^{2}]Y_{LM}(\hat{s}_{i})d\hat{s_{i}}
\end{align}
where $\mathbf{s}_{i}$ is the generate coordinate, $n$ is the number of the
gaussian bases, which is determined by the stability of the results.
By doing this is expansion, we can simplify the integro-differential equation
to an algebraic equation, solve this generalized eigen-equation
to get the energy of the system more easily. The details of solving the RGM
equation can be found in Ref~\cite{RGM}.

\subsubsection{The flavor wave function}
The flavor wave function for the $c\bar{c}s\bar{u}$ system is very simple. For the
meson-meson structure,
\begin{gather}
\chi_{m}^{f1}=[c\bar{c}][s\bar{u}]\\
\chi_{m}^{f2}=[c\bar{u}][s\bar{c}]
\end{gather}
where the superscript of the $\chi$ is the index of the flavor wave function
for meson-meson structure, and the subscript
stands for the isospin $I$ and the third component $I_{z}$.

For the diquark-antidiquark structure,
\begin{equation}
\chi_{d}^{f1}=cs\bar{c}\bar{u}
\end{equation}
The upper and lower indices are similar to those of the meson-meson structure.

\subsubsection{The spin wave function}
For the spin part, the wave functions for two-body clusters are:
\begin{gather}
  \chi^{1}_{\sigma11}=\alpha\alpha
  \qquad
  \chi^{2}_{\sigma10}=\sqrt{\frac{1}{2}}(\alpha\beta+\beta\alpha) \notag\\
  \chi^{3}_{\sigma1 -1}=\beta\beta
  \qquad
  \chi^{4}_{\sigma00}=\sqrt{\frac{1}{2}}(\alpha\beta-\beta\alpha)
\end{gather}
Then, the total spin wave functions for the four-quark system can be obtained
by coupling the wave functions of two clusters.
\begin{align}
\chi^{\sigma 1}_{00}&=\chi^{4}_{\sigma00}\chi^{4}_{\sigma00} \notag\\
\chi^{\sigma 2}_{00}&=\sqrt{\frac{1}{3}}(\chi^{1}_{\sigma11}\chi^{3}_{\sigma1
-1}-\chi^{2}_{\sigma10}\chi^{2}_{\sigma1 0}
+\chi^{3}_{\sigma1-1}\chi^{1}_{\sigma10}) \notag\\
\chi^{\sigma 3}_{11}&=\chi^{4}_{\sigma00}\chi^{1}_{\sigma11} \notag\\
\chi^{\sigma 4}_{11}&=\chi^{1}_{\sigma11}\chi^{4}_{\sigma00} \notag\\
\chi^{\sigma
5}_{11}&=\sqrt{\frac{1}{2}}(\chi^{1}_{\sigma11}\chi^{2}_{\sigma10}-\chi^{2}_{\sigma10}\chi^{1}_{\sigma11})\notag\\
\chi^{\sigma 6}_{22}&=\chi^{1}_{\sigma11}\chi^{1}_{\sigma11}
\end{align}
The spin wave function of two structures is the same.

\subsubsection{The color wave function}
For the meson-meson structure, we give the wave functions for the two-body
clusters first, which are
\begin{gather}
  \chi^{1}_{c[111]}=\sqrt{\frac{1}{3}}(r\bar{r}+g\bar{g}+b\bar{b})\\
  \chi^{2}_{c[21]}=r\bar{b}
  \qquad
  \chi^{3}_{c[21]}=-r\bar{g}\notag\\
  \chi^{4}_{c[21]}=g\bar{b}
  \qquad
  \chi^{5}_{c[21]}=-b\bar{g}\notag\\
  \chi^{6}_{c[21]}=g\bar{r}
  \qquad
  \chi^{7}_{c[21]}=b\bar{r}\notag\\
  \chi^{8}_{c[21]}=\sqrt{\frac{1}{2}}(r\bar{r}-g\bar{g})\notag\\
  \chi^{9}_{c[21]}=\sqrt{\frac{1}{6}}(-r\bar{r}-g\bar{g}+2b\bar{b})
\end{gather}
where the subscript $[111]$ and $[21]$ stand for the color singlet and color
octet cluster respectively.

Then, the total color wave functions for the four-quark system with the
meson-meson structure can be obtained by
coupling the wave functions of two clusters.
\begin{align}
\chi^{c1}_{m}&=\chi^{1}_{c[111]}\chi^{1}_{c[111]}\\
\chi^{c2}_{m}&=\sqrt{\frac{1}{8}}(\chi^{2}_{c[21]}\chi^{7}_{c[21]}-\chi^{4}_{c[21]}\chi^{5}_{c[21]}-\chi^{3}_{c[21]}\chi^{6}_{c[21]}\notag\\
                              {}&+\chi^{8}_{c[21]}\chi^{8}_{c[21]}-\chi^{6}_{c[21]}\chi^{3}_{c[21]}+\chi^{9}_{c[21]}\chi^{9}_{c[21]}\notag\\
                              {}&-\chi^{5}_{c[21]}\chi^{4}_{c[21]}+\chi^{7}_{c[21]}\chi^{2}_{c[21]})
\end{align}
where $\chi^{c1}_{m}$ and $\chi^{c2}_{m}$ represent the color wave function for
the color-singlet channel ($1\times1$) and
the hidden-color channel ($8\times8$), respectively.

For the diquark-antidiquark structure, we firstly give the color wave functions
of the diquark clusters,
\begin{gather}
  \chi^{1}_{c[2]}=rr
  \qquad
  \chi^{2}_{c[2]}=\sqrt{\frac{1}{2}}(rg+gr)
  \qquad
  \chi^{3}_{c[2]}=gg\notag\\
  \chi^{4}_{c[2]}=\sqrt{\frac{1}{2}}(rb+br)
  \qquad
  \chi^{5}_{c[2]}=\sqrt{\frac{1}{2}}(gb+bg)\notag\\
  \chi^{6}_{c[2]}=bb
  \qquad
  \chi^{7}_{c[11]}=\sqrt{\frac{1}{2}}(rg-gr)\notag\\
  \chi^{8}_{c[11]}=\sqrt{\frac{1}{2}}(rb-br)
    \qquad
  \chi^{9}_{c[11]}=\sqrt{\frac{1}{2}}(gb-bg)
\end{gather}
and the color wave functions of the antidiquark clusters,
\begin{gather}
  \chi^{1}_{c[22]}=\bar{r}\bar{r}
  \qquad
  \chi^{2}_{c[22]}=\sqrt{\frac{1}{2}}(\bar{r}\bar{g}+\bar{g}\bar{r})
  \qquad
  \chi^{3}_{c[22]}=\bar{g}\bar{g}\notag\\
  \chi^{4}_{c[22]}=\sqrt{\frac{1}{2}}(\bar{r}\bar{b}+\bar{b}\bar{r})
  \qquad
  \chi^{5}_{c[22]}=\sqrt{\frac{1}{2}}(\bar{g}\bar{b}+\bar{b}\bar{g})\notag\\
  \chi^{6}_{c[22]}=\bar{b}\bar{b}
  \qquad
  \chi^{7}_{c[211]}=\sqrt{\frac{1}{2}}(\bar{r}\bar{g}-\bar{g}\bar{r})\notag\\
  \chi^{8}_{c[211]}=\sqrt{\frac{1}{2}}(\bar{r}\bar{b}-\bar{b}\bar{r})
    \qquad
  \chi^{9}_{c[211]}=\sqrt{\frac{1}{2}}(\bar{g}\bar{b}-\bar{b}\bar{g})
\end{gather}

After that, the total wave functions for the four-quark system with the
diquark-antidiquark structure are obtained as below,
\begin{align}
\chi^{c1}_{d}&=\sqrt{\frac{1}{6}}[\chi^{1}_{c[2]}\chi^{1}_{c[22]}-\chi^{2}_{c[2]}\chi^{2}_{c[22]}+\chi^{3}_{c[2]}\chi^{3}_{c[22]}\notag\\
               {}&+\chi^{4}_{c[2]}\chi^{4}_{c[22]}-\chi^{5}_{c[2]}\chi^{5}_{c[22]}+\chi^{5}_{c[2]}\chi^{5}_{c[22]}]\\
\chi^{c2}_{d}&=\sqrt{\frac{1}{3}}[\chi^{7}_{c[11]}\chi^{7}_{c[211]}-\chi^{8}_{c[11]}\chi^{8}_{c[211]}+\chi^{9}_{c[11]}\chi^{9}_{c[211]}]
\end{align}

Finally, we can acquire the total wave functions by substituting the wave
functions of the orbital, the spin, the flavor
and the color parts into the Eq.(7) according to the given quantum number of
the system.

\section{Numerical results and discussions}
In this work, we investigate the charged hidden-charm tetraquark systems with strangeness in two structures, meson-meson and diquark-antidiquark. We take into account all the possible quantum numbers for the $S-$wave $cs\bar{c}\bar{u}$ systems, which are $IJ^{P}=\frac{1}{2} 0^{+}$, $\frac{1}{2} 1^{+}$, and $\frac{1}{2} 2^{+}$. For the meson-meson structure, we take into account of two color configurations which are the color singlet-singlet$(1\times1)$ and color octet-octet ($8\times8$) configurations in the ChQM, while in the QDCSM, we only consider the color singlet-singlet$(1\times1)$ configuration, since this model considers the effect of the hidden color channel coupling to some extent~\cite{PRC84064001}. For the diquark-antidiquark structure, two color configurations, antitriplet-triplet ($\bar{3}\times3$) and sextet-antisextet ($6\times\bar{6}$), are considered in both models.

To find any bound states of the $cs\bar{c}\bar{u}$ systems, we carry out a dynamical bound-state calculation. The energies of both the single channel and the channel-coupling calculations are obtained. Table~\ref{tab1} shows the results of the meson-meson structure, and Table~\ref{tab2} shows the results of the diquark-antidiquark structure. In the tables, the column headed with $[\chi^{\sigma_{i}}\chi^{f_{j}}\chi^{c_{k}}]$ denotes the combination in spin, flavor and color degrees of freedom for each channel, respectively. The columns headed with $E_{th}$ denotes the theoretical threshold of each channel and $E_{sc}$ represents the lowest energies in the single channel calculation. For meson-meson structure, the column "Channel" represents the physical contents of the channel. The tetraquarks composed of $c\bar{c}s\bar{u}$ can be combined in two ways: $[c\bar{c}][s\bar{u}]$ and $[c\bar{u}][s\bar{c}]$. In each combination, $E_{cc1}$ and $E_{cc2}$ denote the lowest energies of the coupling of the color-singlet channels and the coupling of all channels (including color-singlet channels and hidden-color channels).  The coupling of the two combinations is shown in $E_{cc3}$. For diquark-antidiquark structure, $E_{cc1}$ and $E_{cc2}$ denote the lowest energies of the coupling of the $6\times\bar{6}$ color configuration and the coupling of all channels (including $6\times\bar{6}$ and $\bar{3}\times3$ color configurations). All the general features of the calculated results are as follows.

\subsection{Meson-meson structure}
For the meson-meson structure, $\eta_{c}K$ and $\eta_{c8}K_{8}$ in Table~\ref{tab1} represent the color singlet-singlet$(1\times1)$ and color octet-octet ($8\times8$) configurations, respectively. From Table~\ref{tab1} we can see that the energies of every single channel are above the corresponding theoretical threshold in both two models. The channel-coupling calculation cannot help too much, and the energies by the channel-coupling are still above the theoretical threshold, which indicates that the effect of the channel-coupling is very small here. This is mainly due to the large energy difference between each single channel. As a result, in the meson-meson structure, there is no any bound states in either ChQM or QDCSM. So the reported $Z_{cs}(3985)^{-}$ cannot be explained as $D_{s}D^{*}/D_{s}^{*}D/D_{s}^{*}D^{*}$ molecular state in present calculation.
\begin{table*}[htb]
\begin{center}
\caption{The energies (in MeV) of $c\bar{c}s\bar{u}$ systems with the meson-meson structure in the ChQM and QDCSM.}
\begin{tabular}{ccccccccccc}
  \hline
   \hline
  & & & &\multicolumn{4}{c}{ChQM}& \multicolumn{3}{c}{QDCSM} \\
   \hline
  $IJ^{P}$ & $[\chi^{\sigma_{i}}\chi^{f_{j}}\chi^{c_{k}}] $& Channel & $E_{th}$ & $E_{sc}$ &$E_{cc1}$ & $E_{cc2}$ & $E_{cc3}$&$E_{sc}$ &$E_{cc1}$ & $E_{cc3}$\\
  \hline
  $ \frac{1}{2} 0^{+}$ & $\chi^{\sigma 1}_{00} \chi_{m}^{f1} \chi^{c1}_{m}$ & $\eta_{c}K$ & 3478.2 & 3484.9 & 3484.9 & 3484.9 & 3484.9 & 3484.9 & 3484.9& 3484.9\\
           & $\chi^{\sigma 2}_{00} \chi_{m}^{f1} \chi^{c1}_{m}$ &
           $J/\psi K^{*}$ & 3988.1 & 3994.8 &  & & & 3994.1 &    & \\
           & $\chi^{\sigma 1}_{00} \chi_{m}^{f1} \chi^{c2}_{m}$ & $\eta_{c8}K_{8}$ &  & 4542.6 &  & & & & &\\
           & $\chi^{\sigma 2}_{00} \chi_{m}^{f1} \chi^{c2}_{m}$ & $J/\psi_{8}K^{*}_{8}$ &  & 4363.8 &  &  & & & &\\
           & $\chi^{\sigma 1}_{00} \chi_{m}^{f2} \chi^{c1}_{m}$ & $DD_{s}$ & 3833.4& 3840.6 & 3840.6 & 3838.3 & & 3840.1 & 3840.1&\\
           & $\chi^{\sigma 2}_{00} \chi_{m}^{f2} \chi^{c1}_{m}$ &
           $D^{*}D_{s}^{*}$ & 4119.5 & 4126.2 &  & & & 4121.2 &   &  \\
           & $\chi^{\sigma 1}_{00} \chi_{m}^{f2} \chi^{c2}_{m}$ &
           $D_{8}D_{s8}$ &  & 4395.5 &  & & & & &\\
           & $\chi^{\sigma 2}_{00} \chi_{m}^{f2} \chi^{c2}_{m}$ & $D^{*}_{8}D_{s8}^{*}$ &  & 4125.3 &  & & & & &\\
  \hline
  $\frac{1}{2} 1^{+}$ & $\chi^{\sigma 3}_{11} \chi_{m}^{f1} \chi^{c1}_{m}$ & $\eta_{c}K^{*}$ & 3875.2 & 3881.9 & 3597.8 & 3597.8& 3597.8 & 3881.6 & 3597.8& 3597.7\\
           & $\chi^{\sigma 4}_{11} \chi_{m}^{f1} \chi^{c1}_{m}$ &
           $J/\psi K$ & 3591.1 & 3597.8 &  & & & 3597.8& &\\
           & $\chi^{\sigma 5}_{11} \chi_{m}^{f1} \chi^{c1}_{m}$ &
           $J/\psi K^{*}$ & 3988.1 & 3994.8 &  & & & 3994.1& &\\
           & $\chi^{\sigma 3}_{11} \chi_{m}^{f1} \chi^{c2}_{m}$ & $\eta_{c8}K^{*}_{8}$ &  & 4493.0 &  & & & & &\\
           & $\chi^{\sigma 4}_{11} \chi_{m}^{f1} \chi^{c2}_{m}$ &
           $J/\psi_{8} K_{8}$ &  & 4528.5 &  & & & & &\\
           & $\chi^{\sigma 5}_{11} \chi_{m}^{f1} \chi^{c2}_{m}$ &
           $J/\psi_{8} K^{*}_{8}$ &  & 4423.6 &  & & & & &\\
           & $\chi^{\sigma 3}_{11} \chi_{m}^{f2} \chi^{c1}_{m}$ & $DD_{s}^{*}$ & 3977.7& 3984.4 & 3982.4 & 3980.2 & & 3983.1 & 3981.1 &\\
           & $\chi^{\sigma 4}_{11} \chi_{m}^{f2} \chi^{c1}_{m}$ &
           $D^{*}D_{s}$ & 3975.7& 3982.4 &  & & & 3981.1& &\\
           & $\chi^{\sigma 5}_{11} \chi_{m}^{f2} \chi^{c1}_{m}$ &
           $D^{*}D_{s}^{*}$ & 4119.5 & 4126.2 &  & & & 4121.2& &\\
           & $\chi^{\sigma 3}_{11} \chi_{m}^{f2} \chi^{c2}_{m}$ &
           $D_{8}D_{s8}^{*}$ &  & 4377.5 &  & & & & &\\
           & $\chi^{\sigma 4}_{11} \chi_{m}^{f2} \chi^{c2}_{m}$ &
           $D^{*}_{8}D_{s8}$ &  & 4377.7 &  & & & & &\\
           & $\chi^{\sigma 5}_{11} \chi_{m}^{f2} \chi^{c2}_{m}$ &
           $D^{*}_{8}D_{s8}^{*}$ &  & 4247.2 &  & & & & &\\
  \hline
  $\frac{1}{2} 2^{+}$ & $\chi^{\sigma 6}_{11} \chi_{m}^{f1} \chi^{c1}_{m}$ & $J/\psi K^{*}$ &  3988.1 & 3994.8 & &3994.8  &3994.8 & 3994.1 & &3994.1\\
           & $\chi^{\sigma 6}_{11} \chi_{m}^{f1} \chi^{c2}_{m}$ & $J/\psi_{8} K^{*}_{8}$ &   & 4529.6 &  & &  & & &\\
           & $\chi^{\sigma 6}_{11} \chi_{m}^{f2} \chi^{c1}_{m}$ & $D^{*}D_{s}^{*}$ &  4119.5 & 4126.2 & & 4124.0&  &4121.2  & &\\
           & $\chi^{\sigma 6}_{11} \chi_{m}^{f2} \chi^{c2}_{m}$ &
           $D^{*}_{8}D_{s8}^{*}$ &   & 4462.0 &  &  & & &  &\\
  \hline
\end{tabular}
\label{tab1}
\end{center}
\end{table*}

To study the interaction between two mesons, we carry out the adiabatic calculation of the effective potentials for the $c\bar{c}s\bar{u}$ system. For the color singlet channels, the effective potentials in both two quark models are shown in Fig.~1 and Fig.~2, respectively. In ChQM, all effective potentials are repulsive, and that's why we cannot get any bound state. In QDCSM, although the effective potential of every channel is attractive, it is not strong enough to form any bound state.
For the hidden color channel, we only gives the potential of ChQM in Fig.~3, because the effect of the hidden color channel coupling is considered to some extent in QDCSM.
From Fig.~3 we can see that the minimum potential of each channel appears at the separation of $0.3$ or $0.4$ fm, which indicates that two colorful subclusters are not willing to huddle together or fall apart, so it is possible to form some resonance states here. However, the energy of the hidden color channel listed in Table~\ref{tab1} is among $4.1\sim 4.6$ GeV, which indicates that these resonances may not be suitable to explain the observed $Z_{cs}(3985)^{-}$. The scattering process of the corresponding open channels should be studied to confirm if there is any resonance state or not.
\begin{figure}[h]
\begin{center}
\epsfxsize=3.5in \epsfbox{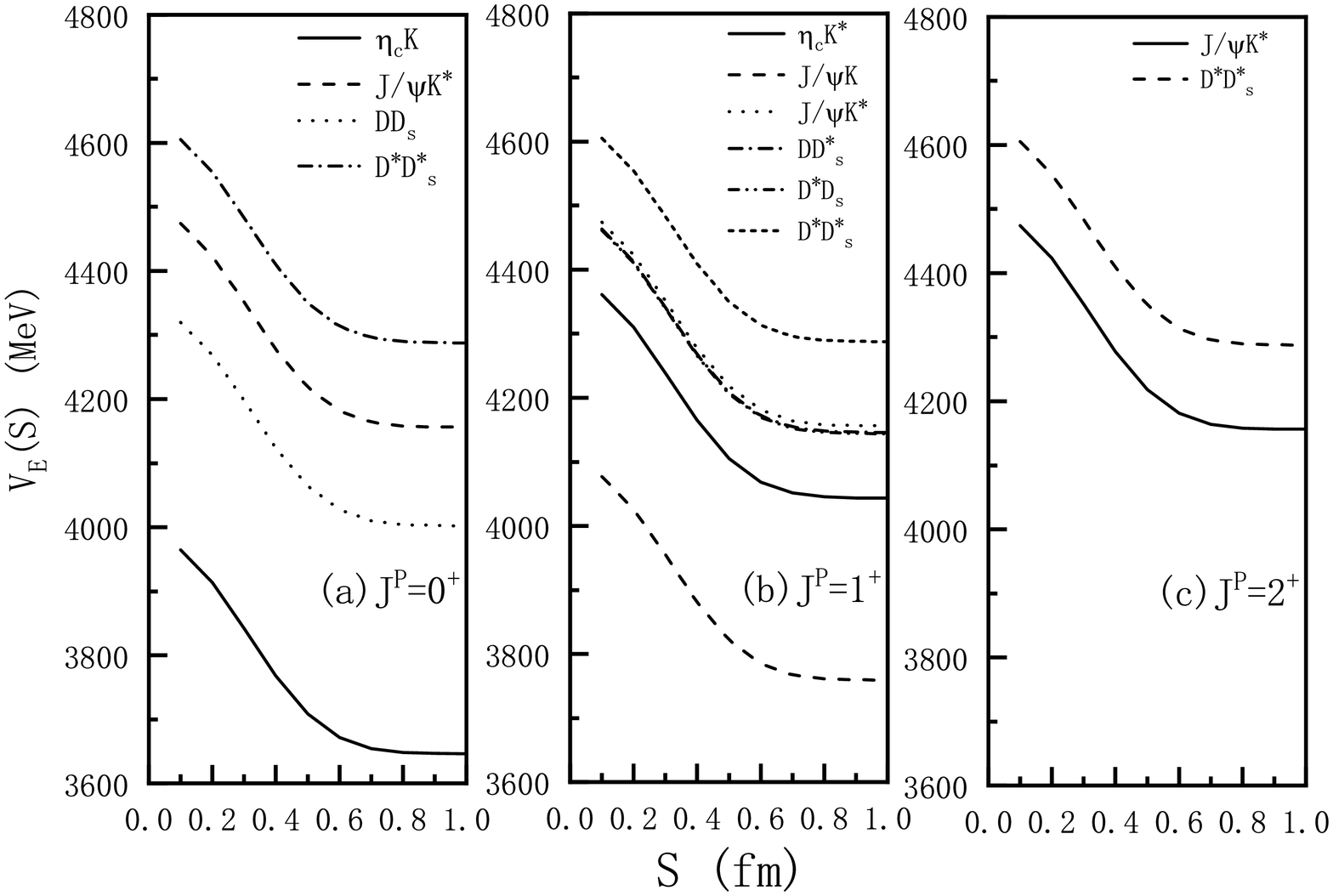} \vspace{-0.3in}

\caption{The effective potentials of the color singlet channels for the meson-meson $c\bar{c}s\bar{u}$ systems in ChQM.} \label{pic1}
\end{center}
\end{figure}
\begin{figure}[h]
\begin{center}
\epsfxsize=3.5in \epsfbox{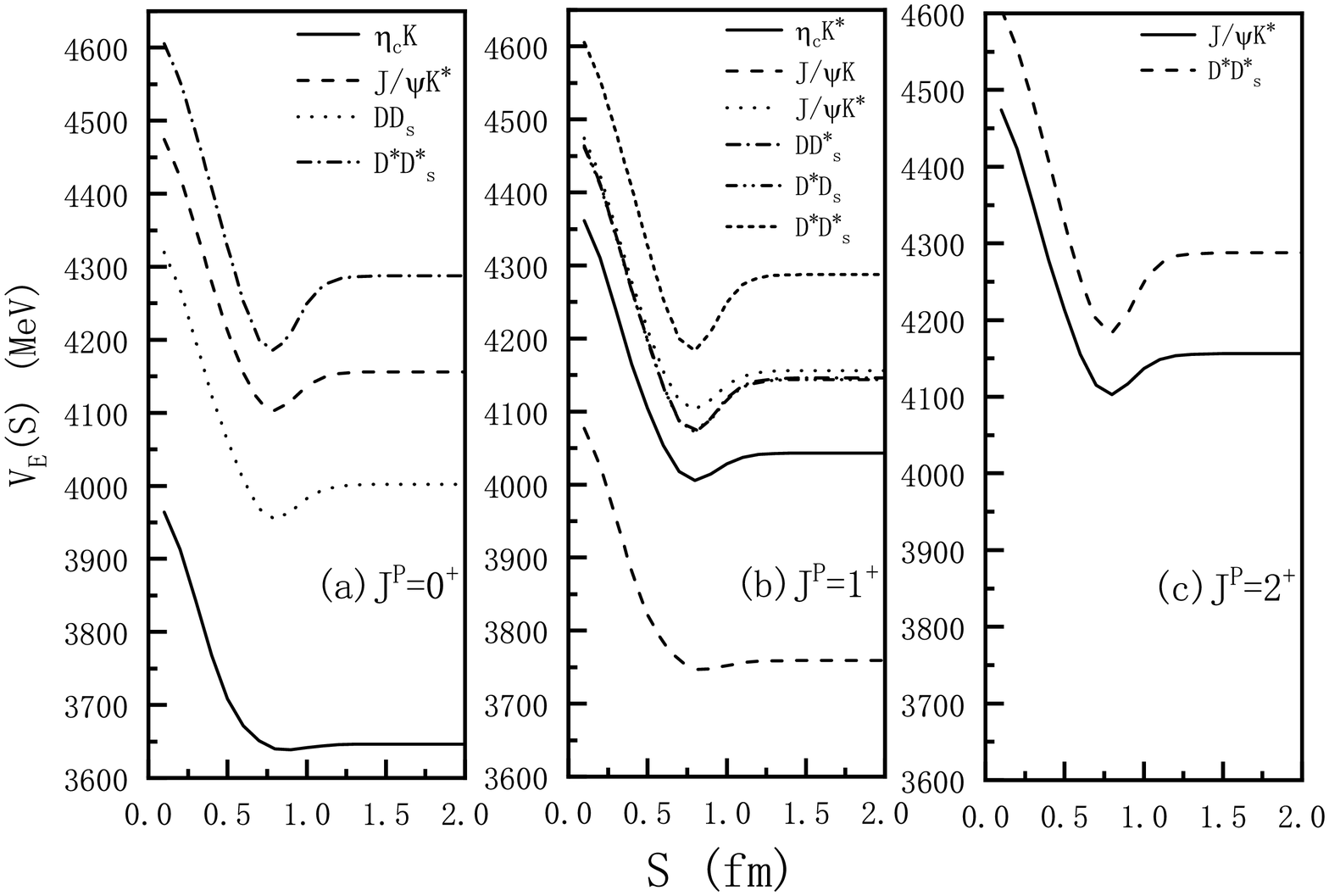} \vspace{-0.3in}

\caption{The effective potentials of the color singlet channels for the meson-meson $c\bar{c}s\bar{u}$ systems in QDCSM.} \label{pic2}
\end{center}
\end{figure}
\begin{figure}[h]
\begin{center}
\epsfxsize=3.5in \epsfbox{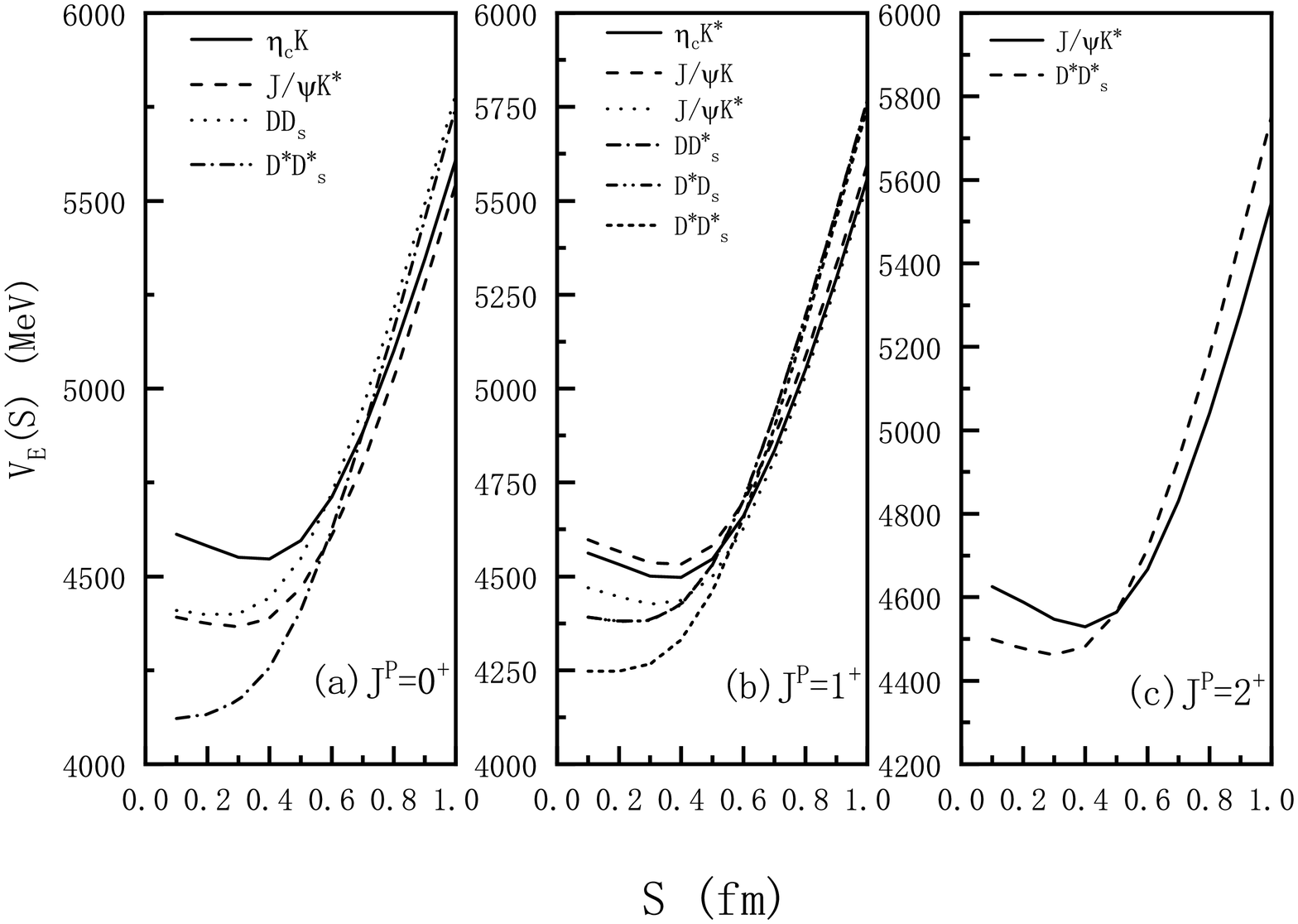} \vspace{-0.3in}

\caption{The effective potentials of the hidden-color channels for the meson-meson $c\bar{c}s\bar{u}$ systems in ChQM.} \label{pic3}
\end{center}
\end{figure}

We also try to investigate why the effective potentials of the two models are different in the color singlet channels. The Hamiltonian of the ChQM and the QDCSM consists of mass, the kinetic energy ($V_{VK}$), the confinement ($V_{CON}$), the Coulomb interaction ($V_{Coul}$) and the color-magnetic interaction ($V_{CMI}$). We take the results of the $IJ^{P}=\frac{1}{2} 0^{+}$ $\eta_{c}K$ channel in meson-meson structure as an example in Fig.~4. We can see that the kinetic energy term in the ChQM provides repulsion, while the one in QDCSM provides attractive interactions. In the ChQM, the confinement ($V_{CON}$), the Coulomb interaction ($V_{Coul}$) and the color-magnetic interaction ($V_{CMI}$) do not contribute to the effective potential between two color singlet clusters $c\bar{c}$ and $s\bar{u}$, so the interaction between the two mesons is only affected by the kinetic energy term($V_{VK}$). There is no exchange terms between $c\bar{c}$ and $s\bar{u}$ in ChQM, so the confinement interaction, the Coulomb interaction and the color-magnetic interaction don't work between $c\bar{c}$ and $s\bar{u}$. Therefore, there is no any term which provides attractive interaction between two color singlet clusters in ChQM, which leads to absence of bound states in this system. In QDCSM, although the quark delocalization is largely weakened because of the heavy quarks in the system, it still affects the kinetic energy, which provides attractive interactions. The other interaction terms are also affected by the quark delocalization and provide repulsive interaction to this system. That's why the total effective potentials are possible to be attractive in QDCSM, although it is still difficult to form any bound state.

\begin{figure}[ht]
\begin{center}
\epsfxsize=3.5in \epsfbox{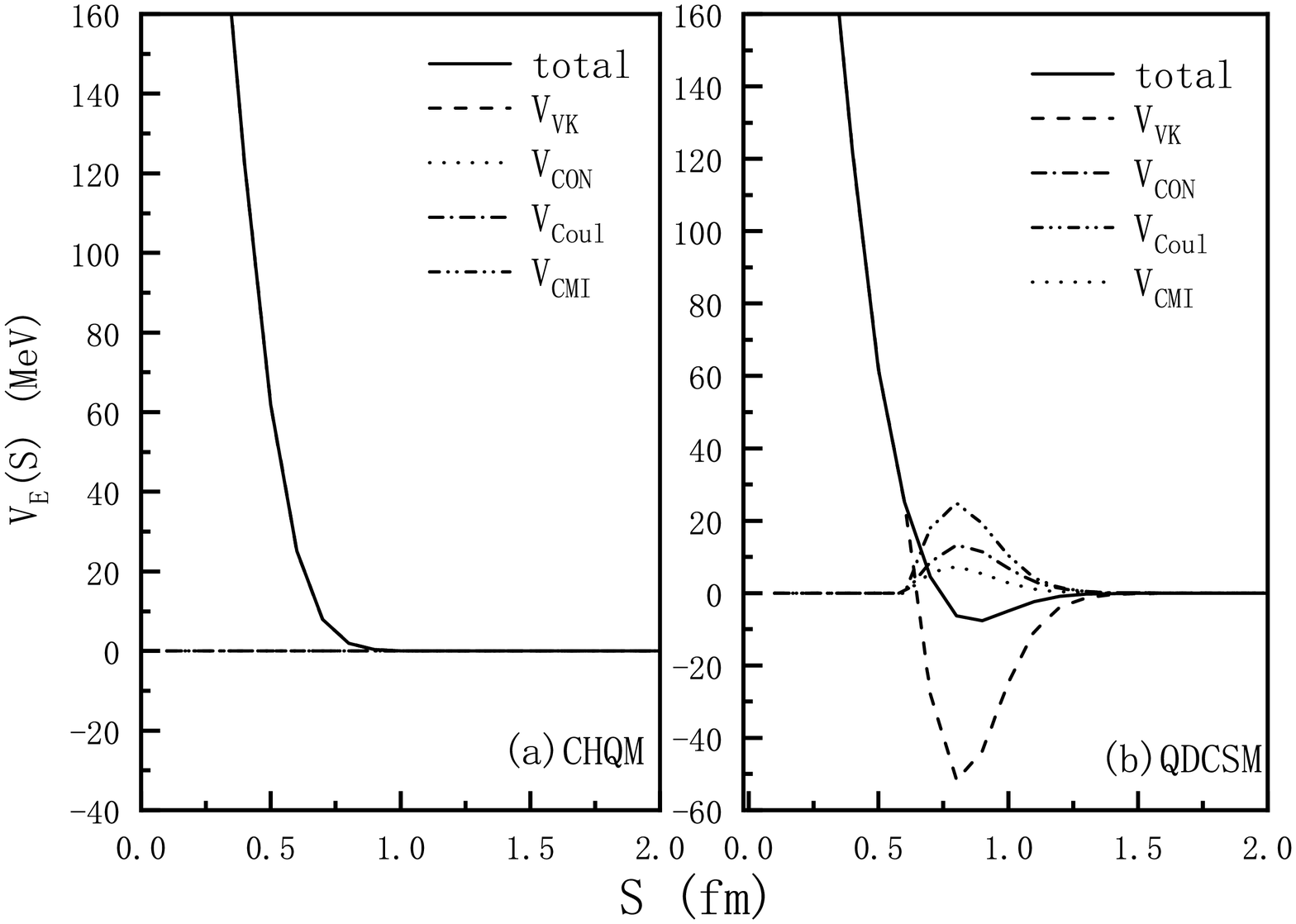} \vspace{-0.3in}

\caption{The contributions to the effective potential of the $IJ^{P}=00^{+}$ $\eta_{c}K$ channel from various terms of interactions in the ChQM and QDCSM.} \label{pic4}
\end{center}
\end{figure}

\subsection{Diquark-antidiquark structure}
\begin{table*}[htb]
\begin{center}
\caption{The energies (in MeV) of $cs\bar{c}\bar{u}$ systems with the diquark-antidiquark structure in the ChQM and QDCSM.}
\begin{tabular}{ccccccccc}
  \hline
   \hline
   & & &\multicolumn{3}{c}{ChQM}& \multicolumn{3}{c}{QDCSM} \\
   \hline
  $IJ^{P}$ & $[\chi^{\sigma_{i}}\chi^{f_{j}}\chi^{c_{k}}] $& $E_{th}$ & $E_{sc}$ &$E_{cc1}$ & $E_{cc2}$ & $E_{sc}$ &$E_{cc1}$ & $E_{cc2}$\\
  \hline
  $ \frac{1}{2} 0^{+}$ & $\chi^{\sigma 1}_{00} \chi_{d}^{f1} \chi^{c1}_{d}$ & 3478.2 & 4462.3 & 4161.1 & 3964.6& 4224.1 & 4034.3 & 3916.5 \\
           & $\chi^{\sigma 2}_{00} \chi_{d}^{f1} \chi^{c1}_{d}$ & &
            4171.6 &  &  & 4041.3 &  &\\
           & $\chi^{\sigma 1}_{00} \chi_{d}^{f1} \chi^{c2}_{d}$ & &
            4263.3 &  &  & 4146.6 &  &\\
           & $\chi^{\sigma 2}_{00} \chi_{d}^{f1} \chi^{c2}_{d}$ & &
            4350.1 &  &  & 4178.5 &  &\\
  \hline
  $\frac{1}{2} 1^{+}$ & $\chi^{\sigma 3}_{11} \chi_{d}^{f1} \chi^{c1}_{d}$  & 3591.1  & 4426.8 & 4245.0 & 4091.2 & 4205.3 & 4089.2 & 4008.8 \\
           & $\chi^{\sigma 4}_{11} \chi_{d}^{f1} \chi^{c1}_{d}$ & &
            4426.3&  &  & 4205.0 &  &\\
           & $\chi^{\sigma 5}_{11} \chi_{d}^{f1} \chi^{c1}_{d}$ & &
            4285.7&  &  & 4116.3 &  &\\
           & $\chi^{\sigma 3}_{11} \chi_{d}^{f1} \chi^{c2}_{d}$ & &
            4334.2&  &  & 4188.3 & & \\
           & $\chi^{\sigma 4}_{11} \chi_{d}^{f1} \chi^{c2}_{d}$ & &
            4335.2&  &  & 4188.8 & & \\
           & $\chi^{\sigma 5}_{11} \chi_{d}^{f1} \chi^{c2}_{d}$ & &
            4379.1&  &  & 4203.6 & & \\
  \hline
  $\frac{1}{2} 2^{+}$ & $\chi^{\sigma 6}_{11} \chi_{d}^{f1} \chi^{c1}_{d}$  &  3988.1 & 4486.5 &  &4418.1 & 4251.7 & & 4246.8\\
           & $\chi^{\sigma 6}_{11} \chi_{d}^{f1} \chi^{c2}_{d}$ & &
            4431.1&  &  &  4251.8 & & \\
  \hline
\end{tabular}
\label{tab2}
\end{center}
\end{table*}

With regards to the diquark-antidiquark structure, the results of both ChQM and QDCSM are listed in Tables ~\ref{tab2}. As shown in the table, the diquark-antidiquark structure has higher energy than the meson-meson structure so there is no any bound state. Besides, the energies in QDCSM are generally lower than
that in ChQM. Since the color symmetry of the diquark and antidiquark are color octet, the color screening will make the quark delocalization
work in QDCSM, which leads to lower energy in this model. We also find that the channel coupling of the diquark-antidiquark structure makes the energy a little lower, but the energy of each state is still higher than the corresponding threshold. However, since the confinement potential requires that the colorful subclusters diquark and antidiquark cannot fall apart directly, resonance states are possible in this configuration. We perform an adiabatic calculation to determine the possibility of the existence of any resonance state, the results of which are shown in Fig. 5.

Obviously, in both ChQM and QDCSM, the energy of each single channel will rise when the two subclusters are too close, so there is a hinderance for the state changing structure to meson-meson even if the energy of the diquark-antidiquark state is higher than the meson-meson state. So it is possible to form a resonance state.
The minimum energy of each channel appears at the separation of $0.3\sim 0.4$fm in ChQM, the one appears at the separation of $0.6\sim0.7$fm in QDCSM, which indicates that the diquark and antidiquark subclusters are close to each other in both two models. Therefore, the resonance state may be the compact resonance state.
According to the Table ~\ref{tab2}, after the channel coupling calculation, the lowest resonance energies in ChQM are $3964.6$ MeV for $IJ^{P}=\frac{1}{2} 0^{+}$, $4091.2$ MeV for $IJ^{P}=\frac{1}{2} 1^{+}$, $4418.1$ MeV for $IJ^{P}=\frac{1}{2} 2^{+}$; and those in QDCSM are $3916.5$ MeV for $IJ^{P}=\frac{1}{2} 0^{+}$, $4008.8$ MeV for $IJ^{P}=\frac{1}{2} 1^{+}$, and $4246.8$ MeV for $IJ^{P}=\frac{1}{2} 2^{+}$. By comparing the mass of the observed $Z_{cs}(3985)^{-}$, this $Z_{cs}$ state is possible to be explained as a compact resonance state $c\bar{c}s\bar{u}$ with $IJ^{P}=\frac{1}{2} 0^{+}$ or $IJ^{P}=\frac{1}{2} 1^{+}$. However, the scattering process of the corresponding open channels should be studied to confirm if there is any resonance state or not.

\begin{figure}[ht]
\begin{center}
\epsfxsize=3.5in \epsfbox{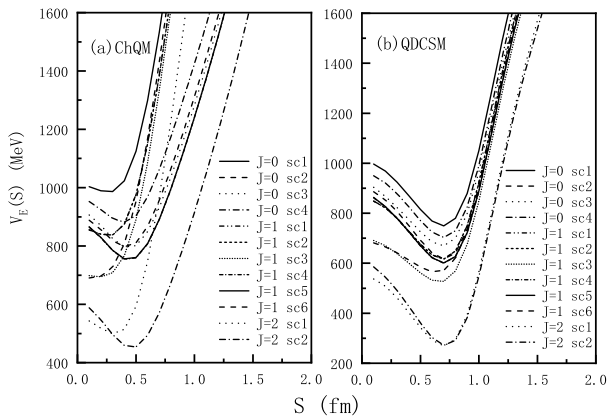} \vspace{-0.3in}

\caption{The effective potentials for the diquark-antidiquark $cs\bar{c}\bar{u}$ systems in two quark models.} \label{pic5}
\end{center}
\end{figure}

\section{Summary}
In this work, we systematically investigate the low-lying charged hidden-charm tetraquark systems with strangeness in two quark models
ChQM and QDCSM. Two configurations, meson-meson and diquark-antidiquark, are considered. The dynamical bound-state calculation is carried out to
search for any bound state in the $cs\bar{c}\bar{u}$ systems. To investigate the effect of the channel coupling, both the single channel and the
channel coupling calculation are performed. Meanwhile, an adiabatic calculation of the effective potentials is added to study the interactions
of the systems and to find any resonance state.

The results are similar in both two quark models. The bound-state calculation shows that there is no any bound state in either ChQM or QDCSM, which excludes the reported $Z_{cs}(3985)^{-}$ as a molecular state $D_{s}D^{*}/D_{s}^{*}D/D_{s}^{*}D^{*}$.
The study of the interaction between two mesons shows that the confinement interaction, the Coulomb interaction and the color-magnetic interaction don't work between two mesons of the $c\bar{c}s\bar{u}$ systems, because there is no exchange terms between them. So it is difficult to obtain a molecular state in this $cs\bar{c}\bar{u}$ system in present work.
However, the effective potentials for the diquark-antidiquark $cs\bar{c}\bar{u}$ systems shows the possibility of some resonance states with mass of $3916.5\sim 3964.6$ MeV for $IJ^{P}=\frac{1}{2} 0^{+}$, $4008.8\sim 4091.2$ MeV for $IJ^{P}=\frac{1}{2} 1^{+}$, $4246.8\sim 4418.1$ MeV for $IJ^{P}=\frac{1}{2} 2^{+}$. The newly reported $Z_{cs}$ state is possible to be explained as a compact resonance state $cs\bar{c}\bar{u}$ with $IJ^{P}=\frac{1}{2} 0^{+}$ or $IJ^{P}=\frac{1}{2} 1^{+}$. To confirm this conclusion, the scattering process of the corresponding open channels is needed, and the decay width can also be obtained by the scattering cross section, which is our further work.

~

\acknowledgments{This work is supported partly by the National Science
Foundation of China under Contract Nos. 11675080, 11775118 and 11535005}.

\end{document}